\documentclass[a4paper]{jpconf}
\usepackage{graphicx}
\begin{document}
\title{Structure of the neutron-rich $N=7$ isotones \\ $^{10}$Li and $^9$He}

\author{H Al Falou\footnote{Present Address: \textsc{TRIUMF}, 4004
Wesbrook Mall, Vancouver, BC V6T 2A3, Canada}, A Leprince and NA Orr}

\address{LPC-ENSICAEN, IN2P3-CNRS et Universit\'{e} de Caen, 14050 Caen cedex, 
France\footnote{Work presented on
behalf of the LPC-CHARISSA-DEMON Collaboration.} }
\ead{alfalou@triumf.ca}

\begin{abstract}
The near threshold structure of the unbound N=7 isotones $^{10}$Li and $^{9}$He 
has been investigated using proton removal and breakup from
intermediate energy (35~MeV/nucleon) secondary beams of $^{11}$Be and $^{14,15}$B.  
The coincident detection of the beam velocity $^{9}$Li and $^{8}$He fragments and
neutrons permitted the relative energy of the in-flight decay of $^{10}$Li and $^{9}$He 
to be reconstructed.  Both systems were found to exhibited virtual $s-$wave strength near threshold together with a higher-lying resonance.
\end{abstract}

\section*{Introduction}

The light nuclei have long provided a test bench for our understanding of
nuclear structure. From an experimental point of view, this is the only region for
which nuclei
lying beyond the neutron dripline are presently accessible. Theoretically, models
incorporating explicitly the continuum are being developed \cite{Michel}. Furthermore, the
structure of
unbound systems, such as $^{10}$Li, is a key ingredient of three-body descriptions
of two-neutron
halo, such as $^{11}$Li, and related nuclei \cite{Jonson}. 

The lightest $N=7$ 
isotones, where the neutron 1$s_{1/2}$ state from the 
1$s$0$d$-shell is found to intrude into the $p$-shell states, are of particular interest.
This phenomenom has long be known in $^{11}$Be \cite{Talmi} and there is now, as cited below, good
evidence that this inversion occurs in $^{10}$Li.  In the case of $^9$He,
experiment suggests that low-lying s-wave strength occurs, although there is
not agreement as to its strength \cite{MSU-He9,GSI-He9}. 
In the following we describe briefly a new experimental 
investigation of the low-lying level structure of $^{10}$Li and $^9$He.

\section*{Experiment}

One of the techniques well suited to the study of nuclei far from stability is that of nucleon removal or breakup of a high-energy radioactive nuclear beam. 
The few-nucleon breakup of such beams can be employed to populate, and study through the
fragment$-$neutron final-state interaction (FSI), unbound nuclei. In
addition to benefiting from significant cross sections
(typically $\sim$10$-$100~mb), the high energies result in the strong forward focussing
of the reaction products (increasing the effective detection acceptances) 
and permit the use of thick targets ($\sim$100~mg/cm$^2$).
Consequently measurements with beam intensities as low as
$\sim$100~pps are feasible.
Here we report 
on measurements using secondary beams 
of $^{11}$Be and $^{14,15}$B to investigate the low-lying level structures of
$^{10}$Li and $^{9}$He.  

The experiments employed 35~MeV/nucleon beams delivered with intensities of some 
10$^4$--10$^5$~pps by the 
\textsc{LISE3} separator at \textsc{GANIL}.  The beam velocity charged fragments and
neutrons emitted in the forward 
direction from the reactions on a carbon target were identified and the 
momenta determined using a Si-Si-CsI array coupled to a large-scale neutron array.
These measurements allowed the fragment+neutron ({\it f-n}) relative energy spectra
to be reconstructed.  
In order to interpret the spectra, simulations, which were validated using the in-flight decay 
of well established
resonances (such as $^{7}$He$_{g.s.}$), were developed to model the response
function of the experimental
setup.  Detailed accounts of the work presented here may be found
elsewhere \cite{Hicham-these,Anne-these}.

\begin{figure}[htb]
 \begin{center}
  \includegraphics[width=5.3in]{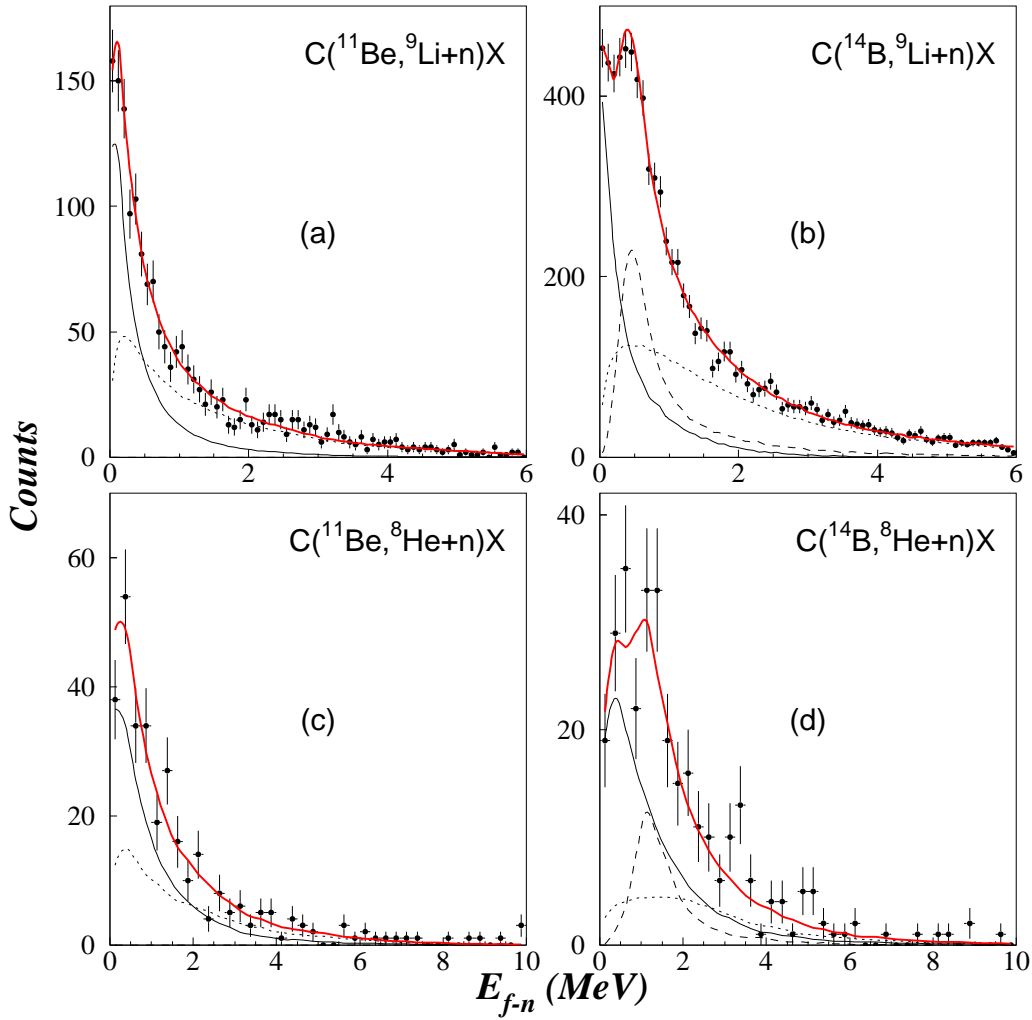}
 \end{center}
\caption{Relative energy spectra for the $^{9}$Li+n and $^{8}$He+n systems for
the different reactions indicated.  
The dotted lines represent the uncorrelated background distribution obtained by event-mixing.  The thin solid
lines are the virtual $s-$states, while the dashed lines are the resonances.  The
thicker
solid line is the overall adjustment.}
\label{fig:dub2}
\end{figure}

\section*{Results}

\begin{figure}[htb]
 \begin{center}
  \includegraphics[height=3.5in]{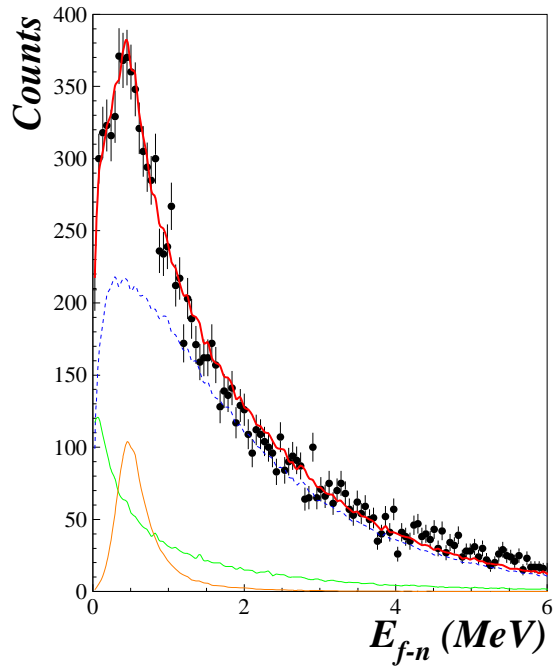}
 \end{center}
\caption{Relative energy spectra for the $^{9}$Li+n from breakup of a $^{15}$B beam.  
The dotted lines represent the uncorrelated background distribution obtained by event-mixing, whilst the thin solid
lines are the virtual $s-$state and resonance (see text).  The thicker
solid line is the overall adjustment.}
\label{fig:anne}
\end{figure}

The results obtained for the $^{8}$He+n and $^{9}$Li+n systems are shown in Figures \ref{fig:dub2} and
\ref{fig:anne}. 
As demonstrated in
our work on the C($^{17}$C,$^{15}$B+n) single-proton removal reaction \cite{JLL-B16}, the description 
of the relative energy spectra
require, in addition to discrete final states, a broad and rather featureless
continuum of 
uncorrelated events which may be generated via event mixing \cite{JLL-these}.  Qualitatively the origin
of these events may be attributed to scattering on the target of the (weakly bound) valence neutron, fragment 
recoil effects \cite{MSU-He9} and
the population of very broad overlapping states.  In the case of breakup involving both proton and neutron removal from the projectile (such as the $^{14,15}$B reactions here), the detection of neutrons arising from the decay of more neutron-rich systems will also contribute.

Both the spectra obtained for the 
$^{9}$Li+n channel (Figures \ref{fig:dub2}(a), \ref{fig:dub2}(b) and \ref{fig:anne}), as well as that for $^{8}$He+n derived from
two-proton removal from $^{11}$Be (Figure \ref{fig:dub2}(c)), exhibit 
significant strength just above threshold, which can be most satisfactorily described by
the presence of a virtual s-wave scattering state.  The results for the
C($^{11}$Be,$^{9}$Li+n) and
C($^{11}$Be,$^{8}$He+n)  reactions
are in line with what may be expected
on the basis of simple considerations, whereby proton only removal from the
projectile should leave the 
neutron configuration undisturbed \cite{MSU-He9,Zinser}.  Given the dominant $s-$wave
neutron component 
in $^{11}$Be$_{g.s.}$,
proton removal to $^{10}$Li and $^{9}$He should populate preferentially $s-$wave
final states.
In the case of $^{9}$Li+n, a scattering length ($a_s$) around
-14~fm  
was deduced, whereas that for $^{8}$He+n is close to 0~fm 
($a_s$=~-3~--~0~fm at the 3-sigma level), signifying a very weak fragment-neutron
interaction.  
The $^{10}$Li result is 
in line with other studies, including high-energy neutron removal from
$^{11}$Li \cite{GSI-Li10-12,GSI-Li10Be13}, whilst that for $^{9}$He, despite being in
some conflict with
very similar work \cite{MSU-He9} to that presented here\footnote{The 
original investigation of the Be($^{11}$Be,$^{8}$He+n) reaction at 25~MeV/nucleon deduced a stronger
FSI, corresponding to a scattering length $a_s$$<$-10~fm \cite{MSU-He9}.}, is in good accord 
with a very recent
report of a study at relativistic energies employing $^{11}$Li breakup \cite{GSI-He9}. 

The $^{9}$Li+n relative energy spectrum from the breakup of 
$^{14}$B (Figure \ref{fig:dub2}(b)) clearly displays the presence of a higher lying state some 0.5~MeV above
threshold, which may be identified with the expected $p-$wave resonance observed in
other studies \cite{GSI-Li10-12,GSI-Li10Be13,Jep07}.
Interestingly, breakup of $^{15}$B exhibits an enhanced yield to this resonance
relative to the 
$s-$state, as displayed in Figure \ref{fig:anne} \cite{Anne-these}.
Despite suffering from limited statistics, the $^{8}$He+n relative energy spectrum
obtained from 
breakup of $^{14}$B (Figure \ref{fig:dub2}(d)) is consistent with the presence of the weakly interacting
$s-$wave 
strength identified above in the two proton-removal from $^{11}$Be and a 
resonance around 1.2~MeV above threshold.  The latter is in line with the original 
observations
made using pion double-charge exchange \cite{Seth} and heavy-ion multi-nucleon 
transfer and reactions \cite{Bohlen,Oertzen}.

\section*{Conclusions}
In summary, in the present work the $\nu s_{1/2}$ character of the $^{10}$Li ground 
state and the existence of a resonance some 0.5~MeV above threshold have been confirmed.  
In addition, evidence for low-lying $s$-wave strength in $^9$He, corresponding to a
rather weak fragment-neutron interaction, has been found.  Indications of a resonance 
some 1.2~MeV above threshold have also been observed.

More generally, the results obtained for $^{10}$Li populated via proton removal from 
the $^{11}$Be beam support the validity at intermediate energies of simple selection rule 
arguments \cite{MSU-He9,Zinser} -- namely, the final-states produced in proton-removal reactions are
dominated by those with the same character as the projectile neutron configuration.  The very weakly bound
nature of $^{11}$Be suggests that such considerations are valid even in the case of removal of
a deeply bound proton from a projectile with a loosely bound valence neuton.

Finally, it was also seen that other final states may be populated in breakup involving proton and 
neutron removal. However, as discussed elsewhere, care must be taken in terms of the 
neutron decay of more neutron-rich systems leading to that of interest \cite{Anne-these,Niigata}. 

\section*{Acknowledgments}
The authors would like to thank their many colleagues in the
\textsc{LPC--CHARISSA--DEMON} Collaboration and
acknowledge the excellent support provided by the technical staff of \textsc{LPC}
and \textsc{GANIL}.


\section*{References}

\begin{thebibliography}{9}

\bibitem{Michel} see, for example, Michel N et al. 2009 {\it J. Phys.} G {\bf 36} 013101 and references therein.

\bibitem{Jonson} see, for example, Jonson B 2004 {\it Phys. Rep.} {\bf 389} 1 and references therein. 

\bibitem{Talmi} Talmi I, Unna I 1960 {\it Phys. Rev. Lett.} {\bf 4} 469;  
Wilkinson DH, Alburger 1959 {\it Phys. Rev. Lett.} {\bf 113} 563.

\bibitem{MSU-He9} Chen L et al. 2001 {\it Phys. Lett.} B {\bf 505} 21.

\bibitem{GSI-He9} Johansson HT, Aksyutina et al. 2010 {\it Nucl. Phys.} A {\bf 842} 15.

\bibitem{Hicham-these} Al Falou H 2007 Ph.D Thesis, Universit\'e de Caen (2007) LPCC
T-07-02;\\
http://tel.archives-ouvertes.fr/docs/00/21/22/14/PDF/thesis.pdf.

\bibitem{Anne-these} Leprince A, Ph.D Thesis, Universit\'e de Caen (2009) LPCC
T-09-04;\\
http://tel.archives-ouvertes.fr/docs/00/45/15/26/PDF/these.pdf

\bibitem{JLL-B16} Lecouey JL et al. 2009 {\it Phys. Lett.} B {\bf 672} (2009) 6.

\bibitem{JLL-these} Lecouey JL 2002 Ph.D Thesis, Universit\'e de Caen  LPCC
T-02-03;\\
http://tel.archives-ouvertes.fr/docs/00/04/54/66/PDF/tel-00003117.pdf.

\bibitem{Zinser} Zinser M et al. 1995 {\it Phys. Rev. Lett.} {\bf 75} 1719.

\bibitem{GSI-Li10-12} Aksyutina Yu et al. 2008 {\it Phys. Lett.} B {\bf 666} 430.

\bibitem{GSI-Li10Be13} Simon H et al. 2007 {\it Nucl. Phys.} A {\bf 791} 267 and
references therein.

\bibitem{Jep07}  Jeppesen H et al. 2006 {\it Phys. Lett.} B {\bf 642} 449 and references therein.

\bibitem{Seth} Seth KK et al. 1987 {\it Phys. Rev. Lett.} {\bf 58} 1930. 

\bibitem{Bohlen} Bohlen H et al. 1988 {\it Z. Phys.} A {\bf 330} 227.

\bibitem{Oertzen} von Oertzen W et al. 1995 {\it Nucl. Phys.} A  {\bf 588} 129c.

\bibitem{Niigata} Al Falou, Leprince A, Orr NA in Proc. of \textsc{NIIGATA2010} -- ``International Symposium on Forefronts of Reseach in Exotic Nuclear Structures'', Tokamachi, Japan, 1-4 March 2010 (in press); 	arXiv:1004.3233v1~[nucl-ex].



\end{thebibliography}
\end{document}